\documentstyle[12pt]{article}
\begin{document}
\centerline{\bf Hong et al. reply to Walliser}
\vskip 1.0cm 
\centerline{ Daniel C. Hong and Paul V. Quinn}
\centerline{ Physics, Lewis Laboratory, 
Lehigh University, Bethlehem, Pennsylvania 18015}
\centerline{Stefan Luding }
\centerline{Particle Technology, Particle Technology,DelftChemTech Julianalaan 136}
\centerline{2628 BL Delft, The Netherlands  }
\vskip 1.0cm
We find it absurd that Walliser [1] essentially used the same analysis and  
obtained identical results as reported in [3], yet arrived at different  
conclusions.   Namely, based on an incomplete theory and using erroneous 
arguments, he  not only disputes the original results [2], but also claims 
them wrong.  A more  complete theory and much more detailed studies were 
published in [3], from  which we concluded that such results support the 
mechanism of segregation  introduced in ref. [2].   We want to make it clear 
that Walliser obtained partial  results of ref. [3] and arrived at the 
opposite conclusion.  In the following we  discuss his comment and its 
relevance, but at the same time point out what  went wrong with his 
arguments.    

First, the free energy functional used in refs. [1,3] is the granular gas/fluid  free  energy.  The observation of either the Brazil-Nut or the reverse Brazil-Nut [2]  was, however, discussed and connected to the condensation or crystallization  of the material at the bottom of the container.  In such very dense situations in  the absence of convection, the geometrical size-segregation presumably  overrides other segregation phenomena [4].  If the system density is nowhere  close to the crystallization density, the fluid free energy description is  appropriate for the granular gas.  The other situations, where parts of the  system are condensed/crystallized, cannot be explained qualitatively by a  gas/fluid free energy - as attempted in the Comment [1] and cannot be  understood by a gas/fluid based approach.  The correct way of describing such  crystallization is to go beyond a simple density functional approach [1,3], and  use the weighted density functional approach, see ref. [5] and references  therein.  Also other attempts based on Enskog theory [6], and/or empirical  predictions for a global equation of state [7] and numerical modeling, at least  account for the crystallization and the corresponding change of material  behavior.   Such advanced methods clearly reveal the formation of crystals below the  condensation temperature. (We want to point out that we used the term  condensation and crystallization interchangeably).   However, the fluid free  energy functional used in [1] and [3] cannot describe the formation of crystals.   Hence, for the mixture of two hard spheres A and B with the condensation  temperatures $T(B)<T(A)$, if the system is quenched between the two  temperatures, $T(B)<T<T(A)$, the method may break down or the results are not  reliable.  This is why the quenching must be done from above.  Nevertheless,  we have considered in [3], contrary to [1], the appearance of the Brazil Nut and  the Reverse Brazil Nut problem for the system quenching $T(B)<T(A)<T$ as a  positive sign to support our condensation driven segregation mechanism.

Second, according to [1], the segregation can be understood by the competition  between gravity and entropy rather
than condensation and percolation.  This  statement is based on a theory, which does not recognize the condensation or  percolation.  Crystallization of hard spheres under gravity is due to the excluded  volume interaction and we have demonstrated analytically [6], numerically [7]  and by Molecular Dynamics simulations [8] that such a hard sphere  crystallization process does exist under gravity.  Furthermore, we have  extended this theory to the binary mixtures in [2,3], and assumed that species  are non-interacting and lead to ideal mixtures.  The scenario was then tested and verified by Molecular Dynamics simulations [2,8].  Therefore, the  formation of the crystal is well grounded and thus nothing controversial has to  be accounted for.  In his comment, Walliser claims that everything can be understood by the conventional thermodynamics.  Without exploring all the  thermodynamic aspects of the segregation, we are not prepared to dispute his argument.  But here are some crucial problems with the thermodynamic  argument.  (i) The stability of the phase diagram obtained in [2].   For a mixture of  large diameter ratio, the phase diagram [2] must breakdown at some  point, and smaller particles on the top must percolate through the  pores and sink to the bottom.  It is questionable whether the  thermodynamics alone can describe such a time dependent stability  problem.  Note that thermodynamics mainly deals with the  equilibrium configurations, and says nothing about the dynamical  process of segregation.    (ii) The correct free energy functional must survive the crystallization, and any conclusions, such as Walliser's [1],  based on pure fluid free energy functional must be incomplete.  It is even dangerous to extract  conclusions from such an incomplete theory.  For a single species,  for example, see ref. [5], where the weighted density functional  theory does yield the crystallization near the bottom of the container.
\vskip 0.2 true cm
\noindent Daniel C. Hong and Paul V. Quinn
\newline
Department of Physics, Lewis Laboratory
\newline
Lehigh University
\newline
Bethlehem, PA 18015  
\vskip 0.2 true cm
\noindent Stefan Luding 
\newline
Particle Technology 
\newline
DelftChemTech Julianalaan 136
\newline
2628 BL Delft, The Netherlands  
\vskip 0.2 true cm
\noindent References  
\vskip 0.2 true cm
\noindent [1] H. Walliser, preceding comment.

\noindent [2] D. C. Hong, P. V. Quinn and S. Luding, Phys Rev. Lett. 86,  3423, (2001).  

\noindent [3] J. A. Both and D. C. Hong, cond-mat/0111347. 

\noindent [4] J. Duran et al, Phys. Rev. Lett.{\bf 70}, 2431 (1993).  

\noindent [5] J. A. Both and D. C. Hong, Phys. Rev. E {\bf 64}, 061105, 
(2001).  

\noindent [6] D. C. Hong, Physica A {\bf 271}, 192 (1999). 

\noindent [7] S. Luding, Phys. Rev. E. {\bf 63}, 042201, (2001).  

\noindent [8] P. V. Quinn and D. C. Hong, Phys. Rev. E. {\bf 62}, 8295 (2000).
\end{document}